\documentclass[10pt,english,conference]{IEEEtran}
\usepackage[T1]{fontenc}
\usepackage[latin9]{luainputenc}
\usepackage{color}
\usepackage{amsmath}
\usepackage{amssymb}
\usepackage{graphicx}

\makeatletter
\usepackage{bm}
\usepackage{tikz}
\usetikzlibrary{shapes,patterns,trees,positioning,matrix,fit,calc,arrows.meta,arrows,decorations.pathreplacing,angles,quotes,shapes.multipart}

\usepackage{pgfplots}
\usepackage{tikzscale}
\usepackage{environ}
\date{}

\@ifundefined{showcaptionsetup}{}{%
 \PassOptionsToPackage{caption=false}{subfig}}
\usepackage{subfig}
\makeatother

\usepackage{babel}
\begin{document}
\title{Iterative Decoding of Trellis-Constrained Codes inspired by Amplitude
Amplification\\
(Preliminary Version)}
\author{\IEEEauthorblockN{Christian Franck} \IEEEauthorblockA{University of Luxembourg\\Computer Science and Communications Research Unit\\6, Avenue de la Fonte, L--4364 Esch-sur-Alzette, Luxembourg\\Email: christian.franck@uni.lu}}
\maketitle
\begin{abstract}
We propose a decoder for Trellis-Constrained Codes, a super-class
of Turbo- and LDPC codes. Inspired by amplitude amplification from
quantum computing, we attempt to amplify the relative likelihood of
the most likely codeword until it stands out from all other codewords.
\end{abstract}

\section{Introduction}

The surprising discovery of Turbo-codes \cite{berrou1993near} in
the early 90's was a major breakthrough in the field of digital communication.
Two simple codes combined with an interleaver can be decoded in a
nearly optimal way with loopy belief-propagation~(BP)~\cite{pearl,mceliceBP}
so that they operate close to Shannon's channel capacity \cite{shannon1949mathematical}.
This lead to the rediscovery of LDPC codes \cite{gallager1962low}
and to the investigation of more general constructions like Trellis-Constrained
Codes (TCCs) \cite{frey1997trellis,franck2016some}. However, it turns
out that near optimal decoding with BP only works for some specific
classes of TCCs, but not in general.

In this paper we describe a method for the probabilistic computation
of the most likely codeword in a TCC w.r.t. a vector of symbol likelihoods.
We iteratively update the symbol likelihoods so that the relative
likelihood of the most likely codeword continually increases until
it hopefully stands out from all other codewords. The algorithm is
insipred by amplitude amplification \cite{brassard1997exact,brassard2002quantum}
which is used in quantum algorithms like Grover search~\cite{grover1996fast}.
Our algorithm converges in a more controlled way than BP.

\begin{figure}
\begin{centering}
\subfloat[A TCC constructed from convolutional codes.]{\begin{centering}
\includegraphics[width=1\columnwidth]{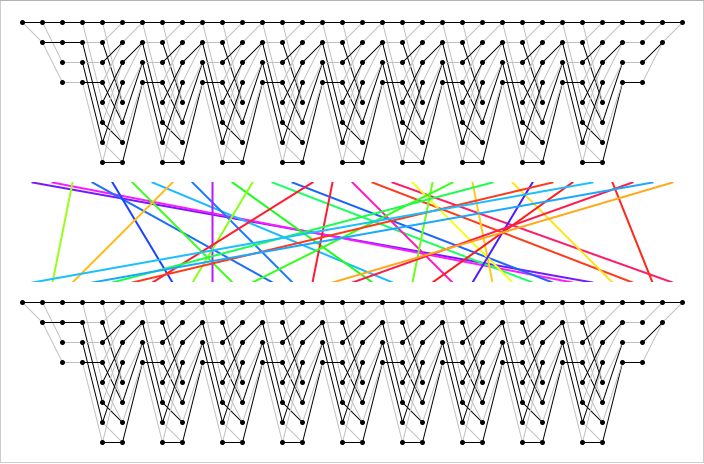}
\par\end{centering}
\label{Fig:TCC-conv}}
\par\end{centering}
\begin{centering}
\subfloat[A TCC representation of a LDPC code.]{\begin{centering}
\includegraphics[width=1\columnwidth]{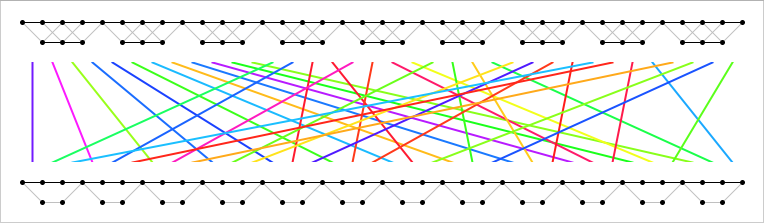}
\par\end{centering}
\label{Fig:TCC-ldpc}}
\par\end{centering}
\caption{Examples of Trellis-Constrained Codes.}
\label{Fig:TCCs}
\end{figure}

\section{Preliminaries}

An intersection code $\mathbb{C}_{\cap}$ is defined as
\[
\mathbb{C}_{\cap}:=\{\mathbf{c}:\mathbf{c}\in\mathbb{C}_{1}\cap\mathbb{C}_{2}\},
\]
where $\mathbb{C}_{1},\mathbb{C}_{2}\subseteq\mathbb{S}=\{-1,+1\}^{n}$
are chosen such that the code $\mathbb{C}_{1}$ and the interleaved
code $\mathbb{C}_{2}$ have a low trellis complexity. Some examples
of TCCs are represented in Fig.~\ref{Fig:TCCs}.

For a memoryless binary channel defined by $\gamma$, a received word
$\mathbf{r}=(r_{1},...,r_{n})\in\mathbb{R}^{n}$ and a word $\mathbf{s}=(s_{1},...,s_{n})\in\mathbb{S}$,
we define the log-likelihood ratio 
\[
L(r):=\frac{1}{2}\ln\frac{P(r|+1)}{P(r|-1)}\text{ with }P(r|s):=\gamma^{rs},
\]
and we use Iverson brackets \cite{knuth1992two}
\[
\langle\mathtt{false}\rangle:=0\text{ and }\langle\mathtt{true}\rangle:=1
\]
to define the code-constrained likelihoods 
\[
P_{\cap}(\mathbf{r}|\mathbf{s}):=\gamma^{\mathbf{r}\mathbf{s}^{T}}\langle\mathbf{s}\in\mathbb{C}_{1}\rangle\langle\mathbf{s}\in\mathbb{C}_{2}\rangle=\gamma^{\mathbf{r}\mathbf{s}^{T}}\langle\mathbf{s}\in\mathbb{C}_{\cap}\rangle.
\]
More details on the channels can be found in the Appendix.

\section{Likelihood Amplification\label{sec:Relative-Likelihood-Amplificatio}}

The objective of an ML decoder is to determine
\[
\check{\mathbf{c}}={\displaystyle \arg\max_{\mathbf{s}\in\mathbb{S}}}\,P_{\cap}(\mathbf{r}|\mathbf{s})={\displaystyle \arg\max_{\mathbf{s}\in\mathbb{C}_{\cap}}}\,\gamma^{\mathbf{r}\mathbf{s}^{T}}.
\]
To reflect the structure of $\mathbb{C}_{\cap}$, with the two contituent
codes and the symbol constraints, we can equivalently write 
\[
(\check{\mathbf{c}},\check{\mathbf{c}})={\displaystyle \arg\max_{(\mathbf{s},\mathbf{s'})\in\mathbb{C}_{1}\times\mathbb{C}_{2}}}\gamma^{\mathbf{w}_{1}\mathbf{s}^{T}+\mathbf{w}_{2}\mathbf{s'^{T}}}\cdot\prod_{j=1}^{n}\langle s_{j}=s'_{j}\rangle
\]
where $\mathbf{w}_{1}+\mathbf{w}_{2}=\mathbf{r}$ with $\mathbf{w}_{1},\mathbf{w}_{2}\in\mathbb{R}^{n}$.

\subsection{Overview}

During the decoding process we iteratively update $\mathbf{w}_{1}$
and $\mathbf{w}_{2}$, and in this description we denote the corresponding
values in iteration $i$ as $\mathbf{w}_{1}^{(i)}$ and $\mathbf{w}_{2}^{(i)}$.
Further we consider
\begin{itemize}
\item the likelihood of the most likely codeword
\[
p_{\check{\mathbf{c}}}^{(i)}:=\gamma^{\mathbf{w}_{1}^{(i)}\check{\mathbf{c}}^{T}+\mathbf{w}_{2}^{(i)}\check{\mathbf{c}}^{T}}\text{, and}
\]
\item the cumulated likelihood of all words in $\mathbb{C}_{1}\times\mathbb{C}_{2}$
\[
\Xi^{(i)}:=\sum_{(\mathbf{s},\mathbf{s'})\in\mathbb{C}_{1}\times\mathbb{C}_{2}}\gamma^{\mathbf{w}_{1}^{(i)}\mathbf{s}^{T}+\mathbf{w}_{2}^{(i)}\mathbf{s}'^{T}}\cdot
\]
\end{itemize}
Initially, we set
\begin{align*}
\mathbf{w}_{1}^{(0)} & \leftarrow\mathbf{r}/2\\
\mathbf{w}_{2}^{(0)} & \leftarrow\mathbf{r}/2
\end{align*}
and we estimate
\[
p_{\check{\mathbf{c}}}^{(0)}=\gamma^{\mathbf{w}_{1}^{(0)}\check{\mathbf{c}}^{T}+\mathbf{w}_{2}^{(0)}\check{\mathbf{c}}^{T}}=\gamma^{\mathbf{r}\check{\mathbf{c}}^{T}}.
\]
Note that for the BEC we have $p_{\check{\mathbf{c}}}^{(0)}=\sum_{i=1}^{n}\langle r_{i}\ne0\rangle$.
Then, in iterations where $i$ is even we compute \begin{align}
\label{eqn:step1} \begin{split}
\mathbf{w}_{1}^{(i+1)} & \leftarrow  \mathbf{w}_{1}^{(i)}+\Delta^{(i)}\\
\mathbf{w}_{2}^{(i+1)} & \leftarrow  \mathbf{w}_{2}^{(i)}-\Delta^{(i)}
\end{split}
\end{align}and in iterations where $i$ is odd we compute\begin{align}
\label{eqn:step2} \begin{split}
\mathbf{w}_{1}^{(i+1)} & \leftarrow \rho^{(i)}\cdot\mathbf{w}_{1}^{(i)}\\
\mathbf{w}_{2}^{(i+1)} & \leftarrow \rho^{(i)}\cdot\mathbf{w}_{2}^{(i)}.
\end{split}
\end{align}The corresponding $\Delta^{(i)}\in\mathbb{R}^{n}$ and $\rho^{(i)}\in\mathbb{R}$
are chosen so that
\begin{equation}
\frac{p_{\check{\mathbf{c}}}^{(i+1)}}{\Xi^{(i+1)}}\geq\frac{p_{\check{\mathbf{c}}}^{(i)}}{\Xi^{(i)}},\label{eq:increasing}
\end{equation}
which means that the relative likelihood of the most likely codeword
stays the same or increases in every step. Details and stopping criteria
are expained in the following sections.

\subsection{Choice of $\Delta^{(i)}$}

In order to chose $\Delta^{(i)}$ so that (\ref{eq:increasing}) holds,
let us investigate the relation between $p_{\check{\mathbf{c}}}^{(i+1)}$
and $p_{\check{\mathbf{c}}}^{(i)}$, and between $\Xi^{(i+1)}$ and
$\Xi^{(i)}$ in (\ref{eqn:step1}).

First, we have 
\begin{equation}
p_{\mathbf{\hat{\mathbf{c}}}}^{(i+1)}=\gamma^{(\mathbf{w}_{1}^{(i)}+\Delta^{(i)}+\mathbf{w}_{2}^{(i)}-\Delta^{(i)})\check{\mathbf{c}}^{T}}=p_{\mathbf{\hat{\mathbf{c}}}}^{(i)}\label{eq:delta1}
\end{equation}
for any $\Delta^{(i)}\in\mathbb{R}^{n}$. The same holds for all
codewords in $\mathbb{C}_{\cap}$, so that the most likely word in
$\mathbb{C}_{\cap}$ under $\mathbf{r}$ also remains the most likely
word under $\mathbf{w}_{1}$ and $\mathbf{w}_{2}$.

\begin{figure}
\begin{centering}
\subfloat[iteration $i$]{\begin{centering}
\begin{tikzpicture}
  \draw[->] (0,0) -- (4,0);
  \draw[->] (0,0) -- (0,2);

  \draw (1,0)[line width=3pt] -- (1,1.00);
  \draw (2,0)[line width=3pt] -- (2,1.75);
  \draw (3,0)[line width=3pt] -- (3,0.5);

\draw (1,0) node[anchor=north] {$\Xi^{(i)}_{-1}$};
\draw (2,0) node[anchor=north] {$\Xi^{(i)}_{0}$};
\draw (3,0) node[anchor=north] {$\Xi^{(i)}_{+1}$};

% \draw (0,0) circle (1cm);
%  \draw (0,0) rectangle (0.5,0.5);
%  \draw (-0.5,-0.5) rectangle (-1,-1);
\end{tikzpicture}
\par\end{centering}
}\subfloat[iteration $i+1$]{\begin{centering}
\begin{tikzpicture}
  \draw[->] (0,0) -- (4,0);
  \draw[->] (0,0) -- (0,2);

  \draw (1,0)[line width=3pt] -- (1,0.60);
  \draw (2,0)[line width=3pt] -- (2,1.75);
  \draw (3,0)[line width=3pt] -- (3,0.60);

\draw (1,0) node[anchor=north] {$\Xi^{(i+1)}_{-1}$};
\draw (2,0) node[anchor=north] {$\Xi^{(i+1)}_{0}$};
\draw (3,0) node[anchor=north] {$\Xi^{(i+1)}_{+1}$};

% \draw (0,0) circle (1cm);
%  \draw (0,0) rectangle (0.5,0.5);
%  \draw (-0.5,-0.5) rectangle (-1,-1);
\end{tikzpicture}
\par\end{centering}
}
\par\end{centering}
\centering{}\caption{Exemplary relation between the values $\Xi_{-1},\Xi_{0},\Xi_{+1}$
in iteration $i$ and $i+1$, for $\Delta^{(i)}=(0,...,\delta_{j\min},...,0)$
with (\ref{eq:deltamin}). It always holds that $\Xi_{-1}^{(i+1)}=\Xi_{+1}^{(i+1)}$
and $\Xi^{(i+1)}\protect\leq\Xi^{(i)}$.}
\label{Fig:Optimization}
\end{figure}
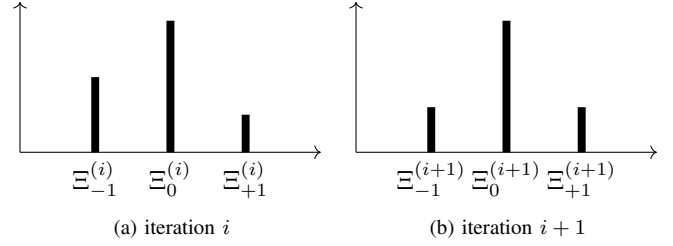

Then, to understand the relation between $\Xi^{(i)}$ and $\Xi^{(i+1)}$
let us assume
\[
\Delta^{(i)}=(0,...,0,\delta_{j},0,...,0)
\]
 with a single possibly non zero value $\delta_{j}$ at position $j$
and
\[
\Xi^{(i)}=\Xi_{-1}^{(i)}+\Xi_{0}^{(i)}+\Xi_{+1}^{(i)}
\]
where
\begin{align*}
 & \Xi_{-1}^{(i)}:=\sum_{(\mathbf{s},\mathbf{s'})\in\mathbb{C}_{1}\times\mathbb{C}_{2}}\gamma^{\mathbf{w}_{1}^{(i)}\mathbf{s}^{T}+\mathbf{w}_{2}^{(i)}\mathbf{s}'^{T}}\cdot\langle s_{j}=-1\rangle\cdot\langle s'_{j}=+1\rangle,\\
 & \Xi_{0{\color{white}+}}^{(i)}:=\sum_{(\mathbf{s},\mathbf{s'})\in\mathbb{C}_{1}\times\mathbb{C}_{2}}\gamma^{\mathbf{w}_{1}^{(i)}\mathbf{s}^{T}+\mathbf{w}_{2}^{(i)}\mathbf{s}'^{T}}\cdot\langle s_{j}=s'_{j}\rangle,\text{}\\
 & \Xi_{+1}^{(i)}:=\sum_{(\mathbf{s},\mathbf{s'})\in\mathbb{C}_{1}\times\mathbb{C}_{2}}\gamma^{\mathbf{w}_{1}^{(i)}\mathbf{s}^{T}+\mathbf{w}_{2}^{(i)}\mathbf{s}'^{T}}\cdot\langle s_{j}=+1\rangle\cdot\langle s'_{j}=-1\rangle.
\end{align*}
It follows from (\ref{eqn:step1}) that
\[
\Xi^{(i+1)}=\gamma^{-2\delta_{j}}\cdot\Xi_{-1}^{(i)}+\Xi_{0}^{(i)}+\gamma^{2\delta_{j}}\cdot\Xi_{+1}^{(i)}.
\]
Thus, for $\delta_{j}=0$ we have $\Xi^{(i+1)}=\Xi^{(i)}$, and for
$\delta_{j}$ equal to 
\begin{equation}
\delta_{j\min}=\arg\min_{\delta_{j}}(\Xi^{(i+1)})=(\log_{\gamma}\Xi_{-1}^{(i)}-\log_{\gamma}\Xi_{+1}^{(i)})/4\label{eq:deltamin}
\end{equation}
we obtain a minimal $\Xi^{(i+1)}$ for which 
\begin{equation}
\Xi^{(i+1)}\leq\Xi^{(i)}.\label{eq:delta2}
\end{equation}
Hence, we can pick a position $j$ and compute $\Delta^{(i)}$ so
that (\ref{eq:delta1}) and (\ref{eq:delta2}) hold. This implies
that (\ref{eq:increasing}) must also hold. Like in quantum computing
the decoder does not care whether the symbols $s^{(1)}=s^{(2)}$ are
both~0 or both~1, the relative likelihood of both states are increased.
The effect of a such an optimization is illustrated in Figure~\ref{Fig:Optimization}.

In practice it can be more efficient to compute $\delta_{1},...,\delta_{n}$
for all symbols at once, according to (\ref{eq:deltamin}), and to
use $\Delta^{(i)}=(\kappa\cdot\delta_{1},...,\kappa\cdot\delta_{n})$,
where  $\kappa$ is a scaling factor that is used to prevent a too
big step that could result from correlations.

\subsection{Choice of $\rho^{(i)}$}

In order to chose $\rho^{(i)}$ so that (\ref{eq:increasing}) holds,
let us investigate the relation between $p_{\check{\mathbf{c}}}^{(i+1)}$
and $p_{\check{\mathbf{c}}}^{(i)}$, and between $\Xi^{(i+1)}$ and
$\Xi^{(i)}$ in (\ref{eqn:step2}).

First, we have
\[
p_{\mathbf{\hat{\mathbf{c}}}}^{(i+1)}=\gamma^{(\rho\cdot\mathbf{w}_{1}^{(i)}+\rho\cdot\mathbf{w}_{2}^{(i)})\check{\mathbf{c}}^{T}}=(p_{\mathbf{\hat{\mathbf{c}}}}^{(i)})^{\rho}.
\]
The same holds for all words in $\mathbb{C}_{\cap}$ and as exponentiation
is monotonous, the most likely codeword for $\mathbf{w}_{1},\mathbf{w}_{2}$
remains also the most likely codeword for $\rho\cdot\mathbf{w}_{1},\rho\cdot\mathbf{w}_{2}$.

Concerning the relation between $\Xi^{(i+1)}$ and $\Xi^{(i)}$, it
is obvious that for $\rho=1$ we have $\Xi^{(i+1)}=\Xi^{(i)}$, but
there is no simple expression for $\rho\ne1$. However, as one can
always compute $\Xi^{(i+1)}$ for a given $\rho$, one can try to
optimize $\rho$ using e.g. a gradient technique. Most importantly,
one can always ensure that (\ref{eq:increasing}) holds, by computing
$p_{\mathbf{\hat{\mathbf{c}}}}^{(i+1)}$ and $\Xi^{(i+1)}$ for a
given $\rho$. We propose and investigate two simple approaches in
the context of our experiments in Section~\ref{sec:Experiments}.

\subsection{Stopping Criteria}

Decoding is successful when $\hat{\mathbf{c}}^{(i)}=(\hat{c}_{1}^{(i)},...,\hat{c}_{n}^{(i)})$
with
\[
\hat{c}_{j}^{(i)}:=\text{sign}\left(\frac{\sum_{(\mathbf{s},\mathbf{s'})\in\mathbb{C}_{1}\times\mathbb{C}_{2}}\gamma^{\mathbf{w}_{1}^{(i)}\mathbf{s}^{T}+\mathbf{w}_{2}^{(i)}\mathbf{s}'^{T}}\langle s_{j}=s'_{j}=+1\rangle}{\sum_{(\mathbf{s},\mathbf{s'})\in\mathbb{C}_{1}\times\mathbb{C}_{2}}\gamma^{\mathbf{w}_{1}^{(i)}\mathbf{s}^{T}+\mathbf{w}_{2}^{(i)}\mathbf{s}'^{T}}\langle s_{j}=s'_{j}=-1\rangle}\right)
\]
is contained in $\mathbb{C}_{1}$ and $\mathbb{C}_{2}$.

\section{Experiments\label{sec:Experiments}}

TBD

\section{Conclusions}

TBD

\bibliographystyle{plain}
\bibliography{refs}

\appendix
We consider channels for which $P(r|s)\propto\gamma^{sr}$, where
the value $\gamma$ is dependent on the channel. For example,
\begin{itemize}
\item the Binary Symmetric Channel (BSC) where
\[
P(r|s):=\begin{cases}
1-p & \text{if }r=s\\
p & \text{if }r\ne s
\end{cases}
\]
so that
\begin{eqnarray*}
L(r) & = & \langle r=+1\rangle\cdot\frac{1}{2}\ln\frac{1-p}{p}+\\
 &  & \langle r=-1\rangle\cdot\frac{1}{2}\ln\frac{p}{1-p}\\
 & = & \left(\frac{1}{2}\ln\frac{1-p}{p}\right)\cdot r
\end{eqnarray*}
and 
\[
\gamma_{\text{BSC}}=\exp\left(\frac{1}{2}\ln\frac{1-p}{p}\right)=\sqrt{(1-p)/p};
\]
\item the Binary Erasure Channel (BEC) where
\[
P(r|s):=\begin{cases}
1-p & \text{if }r=s\\
p & \text{if }r=0\\
0 & \text{if }r=-s
\end{cases}
\]
so that
\begin{eqnarray*}
L(r) & = & \langle r=+1\rangle\cdot\frac{1}{2}\ln\frac{1-p}{0}+\\
 &  & \langle r={\color{white}+}0\rangle\cdot\frac{1}{2}\ln\frac{p}{p}+\\
 &  & \langle r=-1\rangle\cdot\frac{1}{2}\ln\frac{0}{1-p}\\
 & = & \infty\cdot r
\end{eqnarray*}
and (with $\infty\cdot0:=0$ and $\infty^{0}:=1$)
\[
\gamma_{BEC}=\exp\left(\infty\right)=\infty\text{; and}
\]
\item the Additive White Gaussian Noise Channel where
\[
P(r|s):=\frac{1}{\sqrt{2\pi\sigma}}\exp\frac{(r-s)^{2}}{2\sigma^{2}}
\]
so that
\begin{eqnarray*}
L(r) & = & \frac{1}{2}\ln\left(\exp\frac{(r-1)^{2}}{2\sigma^{2}}\right)-\frac{1}{2}\ln\left(\exp\frac{(r+1)^{2}}{2\sigma^{2}}\right)\\
 & = & -\sigma^{-2}\cdot r
\end{eqnarray*}
and
\[
\gamma_{AWGN}=\exp\left(-\sigma^{-2}\right).
\]
\end{itemize}

\end{document}